\documentclass[]{spie}  

\usepackage{amsmath,amsfonts,amssymb}
\usepackage{graphicx}
\usepackage[colorlinks=true, allcolors=blue]{hyperref}
\usepackage{float}

\title{Analyzing misalignment tolerances for implicit electric field conjugation }

\author[a, b]{Joshua Liberman}
\author[c, b]{Sebastiaan Y. Haffert}
\author[b]{Jared R. Males}
\author[a, b]{Kian Milani}

\affil[a]{James C. Wyant College of Optical Sciences, University of Arizona,
Meinel Building 1630 E. University Blvd., Tucson, AZ. 85721}
\affil[b]{Steward Observatory, University of Arizona, 933 N Cherry Ave, Tucson, AZ, USA 85719}
\affil[c]{Leiden University, Leiden Observatory, The Netherlands}

\authorinfo{Further author information: (Send correspondence to Joshua Liberman)\\Joshua Liberman: E-mail: jliberman@arizona.edu}

\begin{document} 
\maketitle

\begin{abstract}
High contrast imaging of extrasolar planets and circumstellar disks requires extreme wavefront stability. Such stability can be achieved with active wavefront control (WFC). The next generation of ground- and space-based telescopes will require a robust form of WFC in order to image planets at small inner working angles and extreme flux ratios with respect to the host star. WFC algorithms such as implicit Electric Field Conjugation (iEFC) reduce stellar leakage by minimizing the electric field within a given region of an image, creating a dark hole. iEFC utilizes an empirical approach to sense and remove speckles in the focal plane. While iEFC is empirically calibrated and can handle optical model errors, there are still model assumptions made during the calibration. The performance of iEFC will degrade if the system changes due to slow, optomechanical drifts. In this work, we assess the iEFC performance impacts of pupil misalignments on the deformable mirror (DM) and focal plane misalignments on the detector. We base our analysis on the MagAO-X instrument, an extreme AO system installed on the Magellan-Clay telescope, to develop iEFC misalignment tolerancing requirements for both ground- and space-based missions. We present end-to-end physical optics simulations of the MagAO-X instrument, demonstrating iEFC’s alignment tolerance.  
\end{abstract}


\section{INTRODUCTION}
\label{sec:intro}  

Direct imaging of Earth-like planets in the habitable zone will require extreme wavefront stability and control\cite{Currie2022}. The next generation of ground-based extremely large telescopes (ELTs) will target Earth-like planets around low-mass stars, requiring raw contrasts of $\approx 10^{-5}$ at $\approx 0.015"$\cite{kasper2021, close2022}. 
On the ELTs, meeting the contrast goals necessary for exo-Earth imaging at close separations requires extreme adaptive optics (ExAO) to remove atmospheric residuals and improve resolution as well as coronagraphy to suppress starlight at the focal plane. NASA's Habitable Worlds Observatory--a space-based direct imaging mission recommended by the Astro2020 decadal survey--imposes stricter contrast requirements of $10^{-10}$ at $\approx 0.1"$ separations\cite{astro2020,luvoir2019, habex2020}.

Both ground- and space-based telescopes suffer from non-common path aberrations (NCPAs): wavefront errors originating from within the optical system. NCPAs create quasi-static speckles in the focal plane. These NCPAs are not seen by ground-based ExAO systems. Furthermore, quasi-static speckles occurring at close separations are difficult to remove with post-processing, limiting raw contrast in both ground- and space-based systems\cite{Haffert2023}.

Coronagraphic imaging systems typically consist of a focal plane mask (FPM) to suppress on-axis starlight and a Lyot stop to block light that is diffracted by the FPM. Additionally, a deformable mirror (DM) is used to correct for NCPAs in a process known as higher order wavefront sensing and control (HOWFSC)\cite{redmond2021}.
HOWFSC can suppress quasi-static speckles, enabling deeper contrasts and enhancing signal-to-noise ratio. 

Implicit Electric Field Conjugation (iEFC) is a measurement-based HOWFSC technique, utilizing a Jacobian matrix to map the response between the DM and modulated intensity measurements. This matrix can, in turn, be used to implicitly reconstruct the electric field and derive a DM solution that creates a dark hole (DH) region free of speckles\cite{Haffert2023}. The iEFC calibration is performed with a well-aligned, coronagraphic optical system, however, both ground- and space-based systems are susceptible to slow drifts of the optics. These drifts cause light to leak through the coronagraph and into the DH region\cite{soummer2007}. As a result, contrast performance degrades over time. 

Drifts in the telescope optical system can arise from multiple sources. Thermal gradients may induce deformations in both the optical mounts and the table itself, causing optics to shift\cite{soummer2007}. Additionally, for ground-based instruments mounted at the Nasmyth foci, dynamic pupil tracking misalignments may occur during observation. The Magellan Extreme Adaptive Optics (MagAO-X) instrument is mounted at the Nasmyth port of the Magellan-Clay 6.5m telescope. The Clay Nasmyth port rotates with elevation but the instrument does not. This introduces a clocking misalignment of the MagAO-X masks with respect to the telescope pupil. A K-mirror is used to counter the pupil rotation\cite{hedglen2018}. However, rotation of the K-mirror, in turn, misaligns both the pupil and the focal planes in the instrument, causing slow drifts within observations. We performed end-to-end optical simulations, based on the MagAO-X instrument architecture, to characterize iEFC performance in the presence of optical misalignments.

\section{METHODS}
\begin{figure}[H]
\begin{center}
\begin{tabular}{c}
\includegraphics[height=8cm,]{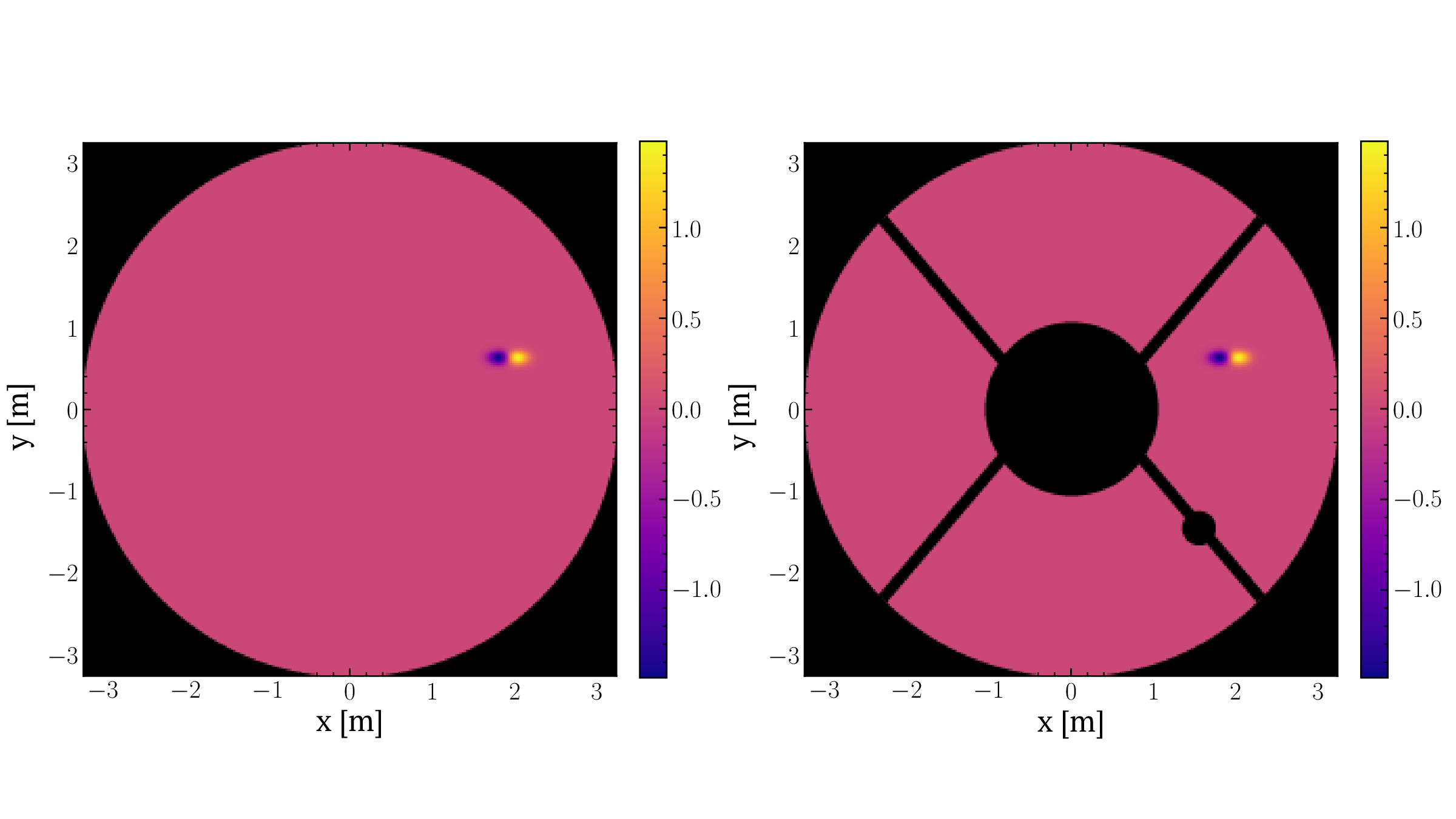}
\end{tabular}
\end{center}
\caption 
{ \label{fig:pre-apod}
The clear aperture (left) and the MagAO-X aperture with the pre-apodizing bump mask (right). Locations of the pairwise probes used in iEFC are denoted with spots.} 
\end{figure}

To assess misalignment tolerancing requirements for iEFC, we developed an end-to-end physical optics simulation based on the current layout of the MagAO-X instrument--an ExAO system that is designed to image exoplanets at visible wavelengths\cite{males2018}. We constructed our simulation with the High Contrast Imaging for Python (HCIPy) library\cite{por2018}. We simulated two different coronagraph designs: a perfect coronagraph and a charge 6 vector vortex coronagraph (VVC)--a variation of the optical vortex coronagraph\cite{mawet2010}. The vortex and perfect coronagraphs make for a good side-by-side comparison, as they behave similarly. The perfect coronagraph suppresses all light for a flat wavefront regardless of aperture. The vortex is close to a perfect coronagraph, diffracting all on-axis light outside of the Lyot stop, for clear apertures\cite{swartzlander2008}. Two different apertures were used in our simulation models. For the perfect coronagraph, we simulated the Magellan 6.5m telescope pupil with a pre-apodizing bump mask to hide an obstruction on the MagAO-X 2k DM (Fig.~\ref{fig:pre-apod})\cite{VanGorkom2021}. For the VVC, we use a clear telescope aperture, as the VVC exhibits sub-optimal performance with a centrally-obscurated pupil\cite{fogarty2017}. We also include a slightly-undersized Lyot stop with a diameter of 95\% of the incoming beam diameter. Note that the perfect coronagraph does not need a Lyot stop, as it suppresses all on-axis light at the focal plane. We perform wavefront control with a 34x34 actuator Boston Micromachines DM to mimic the non-common path DM used on MagAO-X\cite{males2022}. All simulations were computed with a monochromatic source ($\lambda = $750 nm). Phase aberrations are simulated using a power law spectral density with an exponential term of -2.5 and a peak-to-valley wavefront error amplitude of 0.1$\lambda$.

To construct our measurement Jacobian for iEFC, we use pairwise probes to generate modulated intensity images and recover the focal plane intensity. We also use calibration probes to construct our iEFC response matrix. We generate the pairwise probes by applying 2 single actuator pokes with amplitudes of 0.01$\lambda$ to adjacent actuators on the DM (Fig.~\ref{fig:pre-apod}). We generate DM calibration probes from a basis of Fourier modes and an amplitude of 0.01$\lambda$ as before.

We then pseudo-invert our response matrix using Tikhonov Regularization to generate a reconstruction matrix that maps the effect of each Fourier mode onto a corresponding pixel in the post-coronagraphic focal plane image. We adopt a regularization strength of $2*10^{-3}$ relative to the maximum singular value of our response matrix. This regularization parameter remains constant across all simulations.

\begin{figure}[H]
\begin{center}
\begin{tabular}{c}
\includegraphics[height=6.5cm]{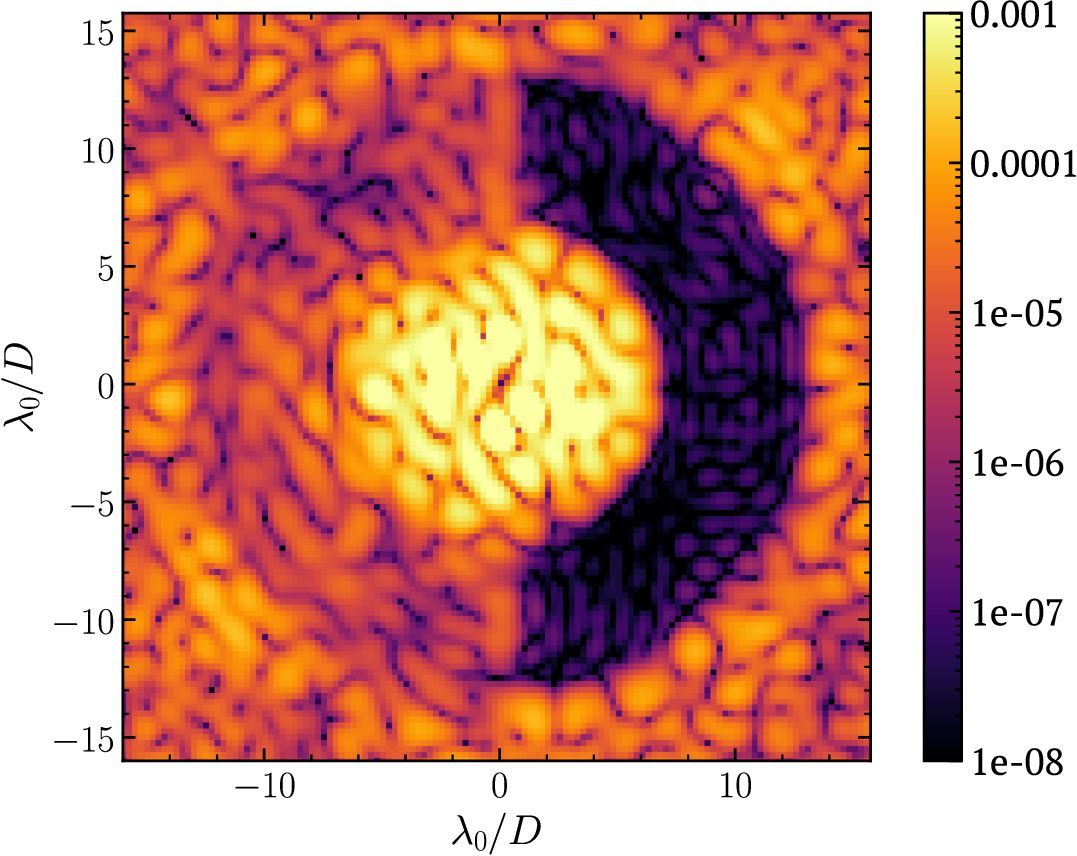}
\end{tabular}
\end{center}
\caption 
{ \label{fig:post-iefc-psf}
A simulated PSF after correction with iEFC (7-13 $\lambda / D$ dark hole).} 
\end{figure}

A simulated stellar PSF with a D-shaped dark hole over 7-13$\lambda / D$ is shown in Fig.~\ref{fig:post-iefc-psf}. Note that, while a single DM creates a one-sided dark hole, iEFC also removes some speckles in the region opposite the dark hole. This behavior is expected, as we only introduced aberrations in phase rather than phase+amplitude.

\section{ANALYSIS}
To analyze iEFC sensitivity to misalignments, we construct a response matrix for a perfectly aligned optical system. We then intentionally misalign an optic and characterize the degradation in contrast for the iEFC solution. We define contrast performance using the normalized intensity (NI) metric, where NI is the average flux within the DH region for the on-axis coronagraphic PSF divided by the maximum number of counts in the non-coronagraphic stellar PSF.

\begin{figure}[H]
\begin{center}
\begin{tabular}{c}
\includegraphics[height=6.5cm]{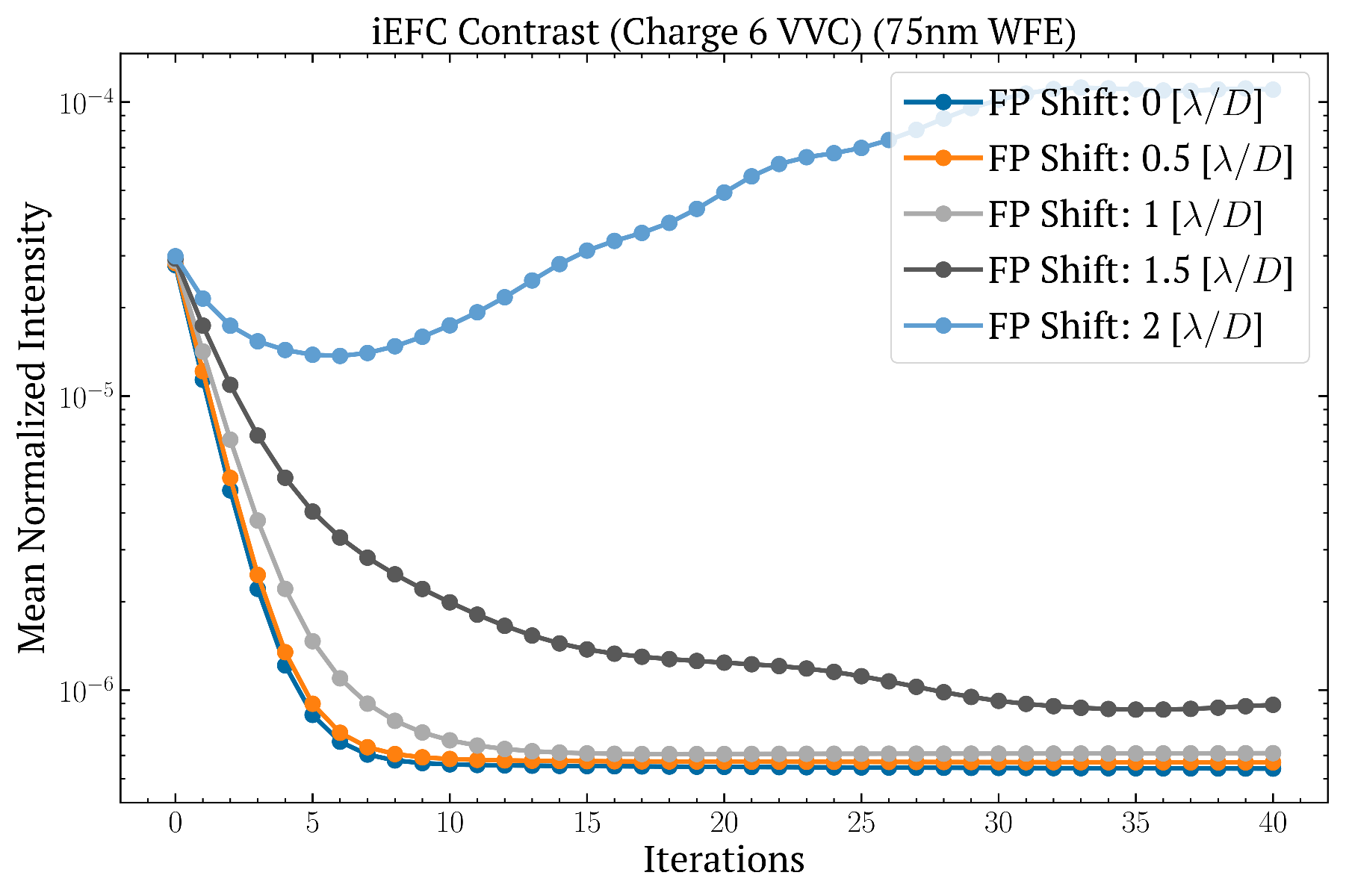}
\end{tabular}
\end{center}
\caption 
{ \label{fig:fpm-offset-vvc}
iEFC contrast for the charge 6 VVC at various FPM offsets on the detector.} 
\end{figure}

Accurate alignment of the coronagraph focal plane mask (FPM) is necessary for maintaining raw instrument contrasts\cite{matthews-2017}. 
In Fig.~\ref{fig:fpm-offset-vvc}, we misalign the FPM with respect to the detector. The iEFC solution converges for shifts of the FPM up to 1.5 $\lambda / D$. Our results are comparable to those of Ref.~\citenum{matthews-2017}, where the authors find that EFC performance degrades for FPM misalignments $\geq$1$\lambda / D$ in simulations of the Palomar Project 1640 coronagraph. The discrepancy in results may be due to a difference in the FPMs used between studies. The authors in Ref.~\citenum{matthews-2017} use an apodized pupil Lyot coronagraph rather than a charge 6 VVC.

\begin{figure}[H]
\begin{center}
\begin{tabular}{c}
\includegraphics[height=6.5cm]{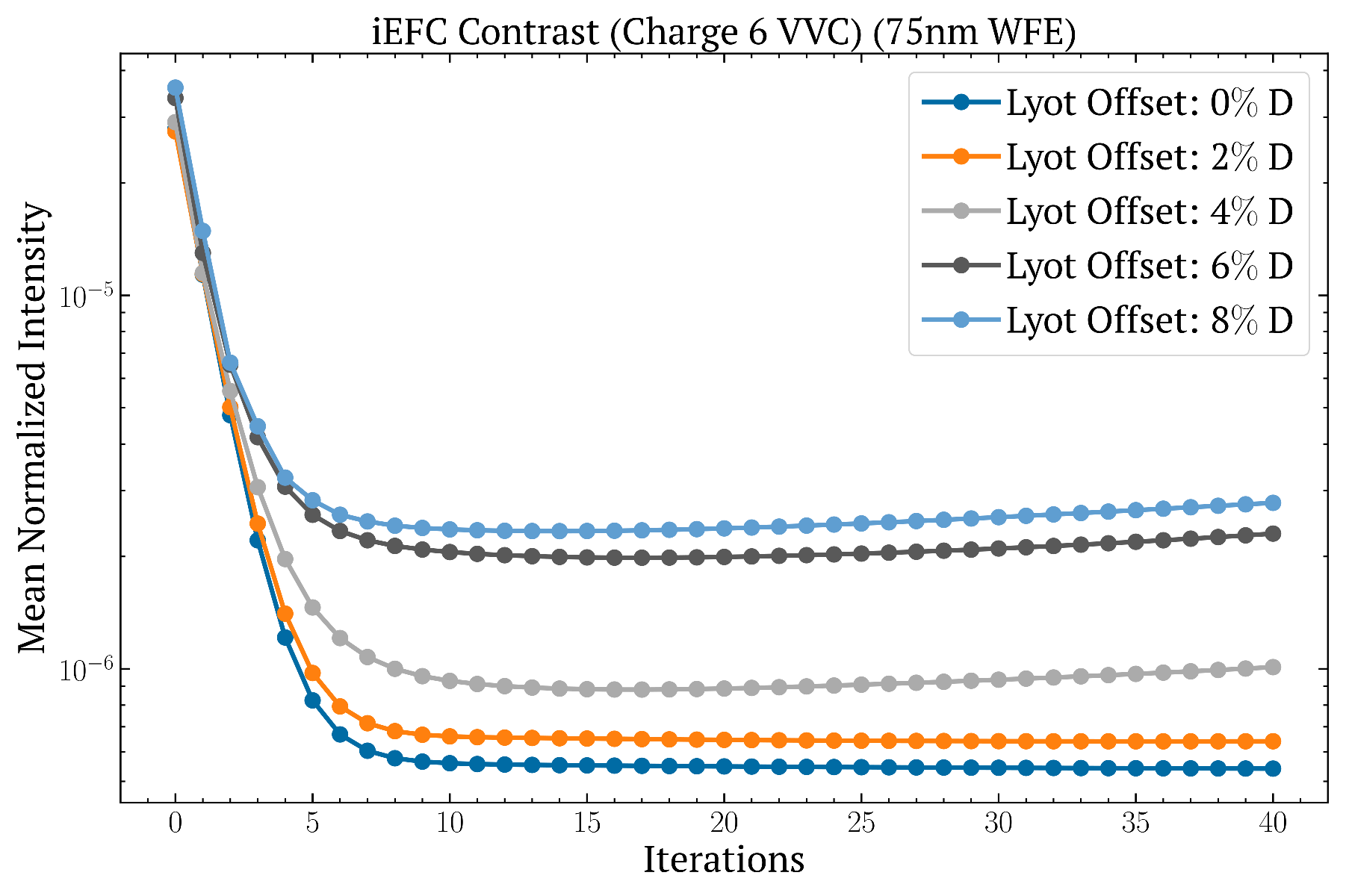}
\end{tabular}
\end{center}
\caption 
{ \label{fig:lyot_offset_vortex}
iEFC response to Lyot stop misalignments for the charge 6 VVC.} 
\end{figure}

The Lyot stop suppresses on-axis starlight diffracted by the VVC. Lyot stop misalignments will increase the stellar intensity in the focal plane, degrading contrast.
In Fig.~\ref{fig:lyot_offset_vortex}, we offset the Lyot stop position for the charge 6 VVC along the x-axis with respect to the telescope pupil. We observe that the iEFC solution is fairly insensitive to misalignments of the Lyot stop, with the contrast degrading at offsets of $\approx 6\% D$. For comparison, the Lyot stop is undersized by $95\% D$. These results demonstrate iEFC's robustness to misalignments of the Lyot stop. We are able to dig dark holes of $\approx 10^{-6}$ contrast despite having some diffracted light leak past the Lyot stop. 

\begin{figure}[H]
\begin{center}
\begin{tabular}{c}
\includegraphics[height=6.5cm]{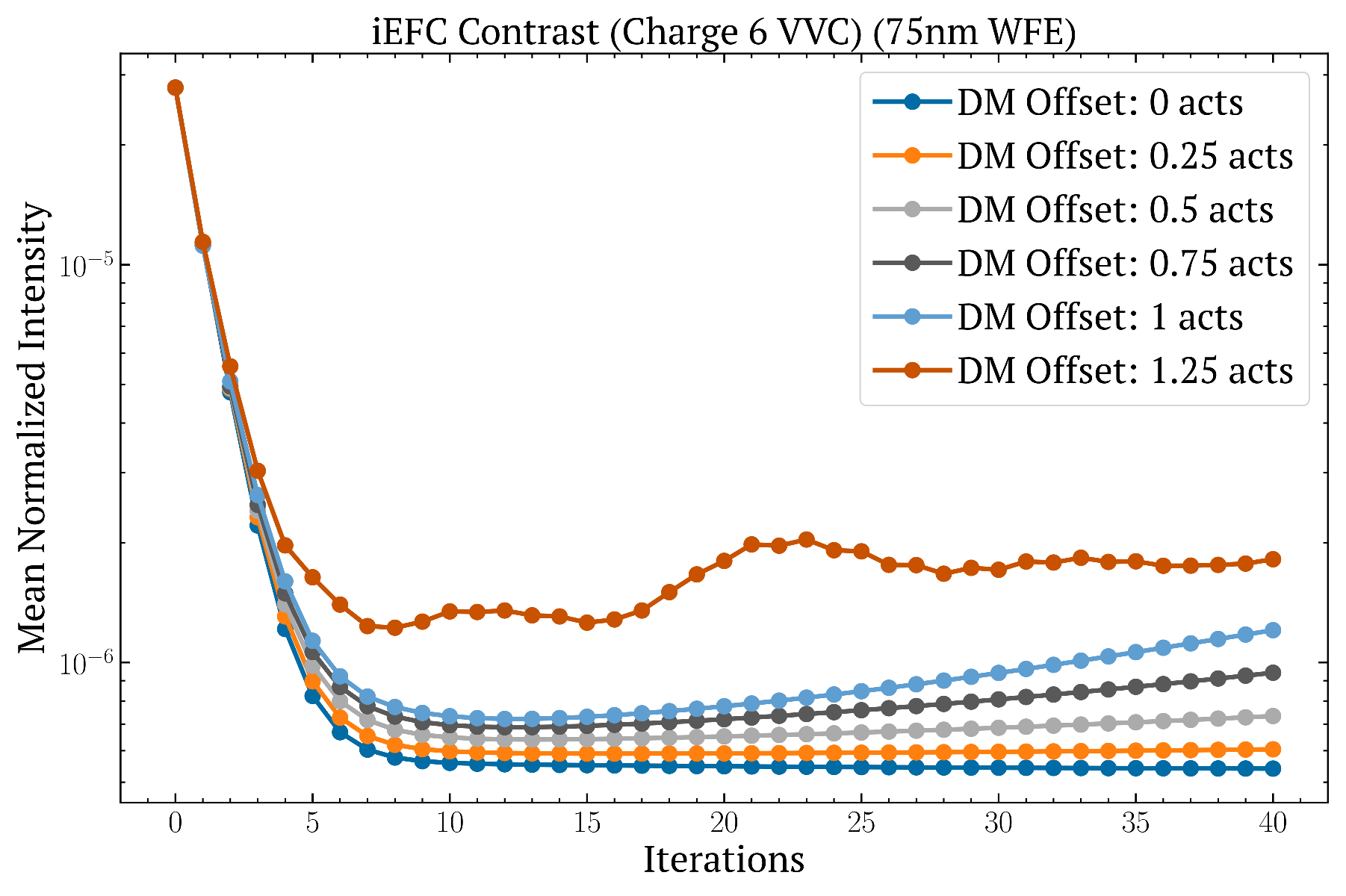}
\end{tabular}
\end{center}
\caption 
{\label{fig:dm-shift-vortex}
iEFC sensitivity to DM misregistration effects for the charge 6 VVC.} 
\end{figure}

Misregistration of the DM negatively impacts the iEFC solution, as the applied phase on the DM will no longer correspond to the phase prescribed by the Jacobian. Instead, the phase will be shifted, transforming sine waves into cosine waves on the DM. In Fig.~\ref{fig:dm-shift-vortex}, we simulate DM registration errors, misaligning the DM in the x-direction with respect to the optical pupil. Simulations of DM misregistration indicate that the 1k DM can be misaligned by $\leq$0.5 actuators before the iEFC solution begins to diverge.

\begin{figure}[H]
\begin{center}
\begin{tabular}{c}
\includegraphics[height=6.5cm]{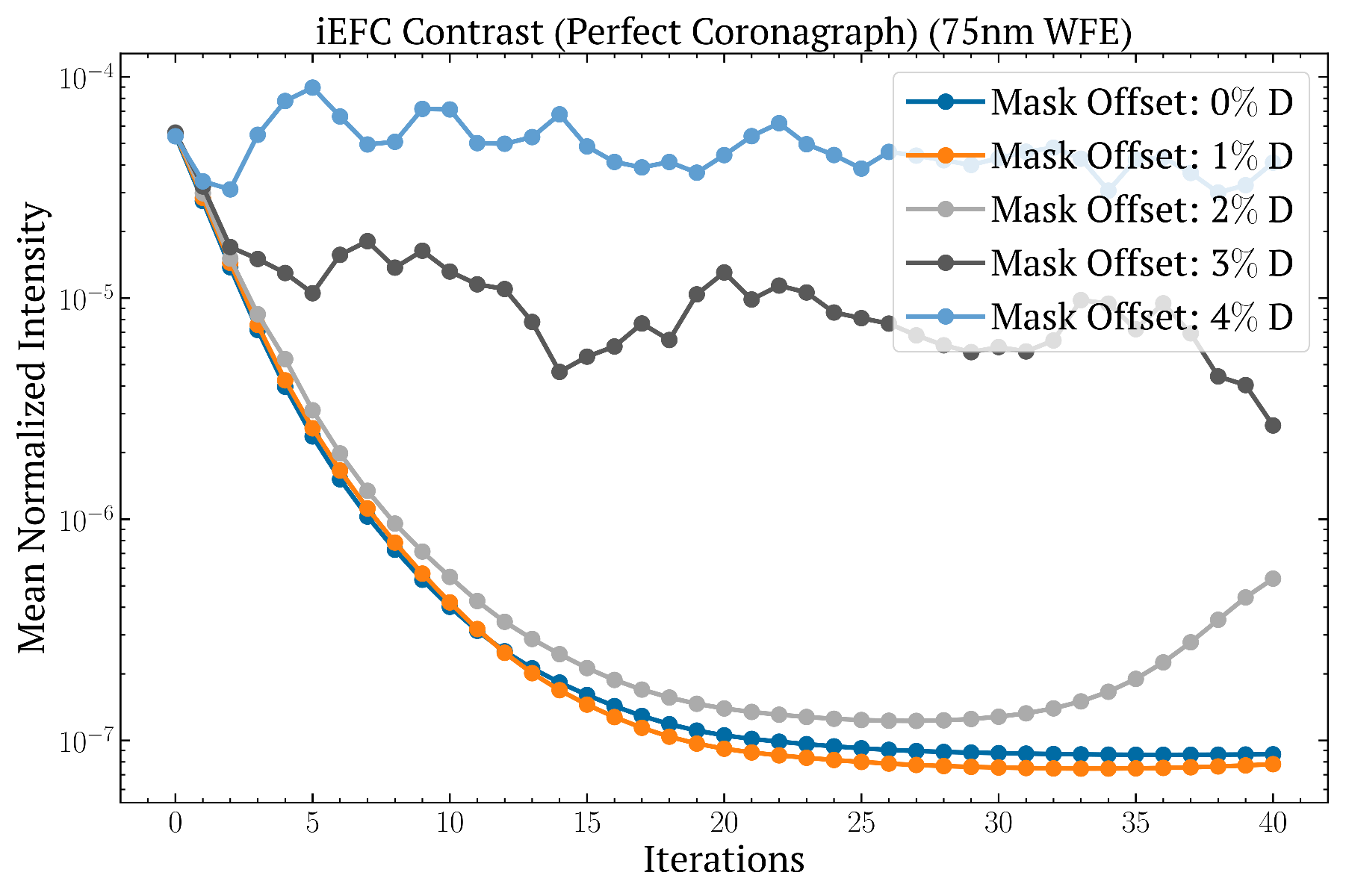}
\end{tabular}
\end{center}
\caption 
{\label{fig:bump-offset}
iEFC response to offsets of the MagAO-X pre-apodizer for the order 2 perfect coronagraph.} 
\end{figure}

Many ground-based extreme AO systems include pre-apodizing masks to hide dead actuators or obstructions on the tweeter DM\cite{nguyen2022, VanGorkom2021, sauvage2016, skaf2022}. MagAO-X contains a pre-apodizing bump mask to hide an obstruction that prevents the 2k tweeter DM surface from achieving full stroke (Fig.~\ref{fig:pre-apod}). In Fig.~\ref{fig:bump-offset}, we offset the MagAO-X bump mask position with respect to the pupil. We find that, beyond an offset of 1$\% D$, the iEFC solution rapidly diverges. iEFC's hypersensitivity to misalignments of the bump mask is not surprising, as subtle shifts in the mask will expose defective actuators or obstructions that do not have a measured response in the Jacobian.

\section{CONCLUSION}
In this work, we measured the impacts of optical misalignments on iEFC post-calibration. We found that iEFC is most sensitive to misalignments of the pre-apodizer--a mask that is used in ground-based ExAO systems to hide dead actuators on the pupil, while iEFC is relatively insensitive to misalignments of the Lyot stop. We determine that developing an optical tolerancing budget for iEFC allows for relaxed telescope stability requirements in ground- and space-based imaging missions. 

Future work will involve expanding the tolerancing simulations to include more coronagraph designs, such as the knife-edge and Lyot architectures. We will also examine how the iEFC solution is affected by variable spectral energy distributions (SEDs), as the on-sky target will likely have an SED that differs from that of the calibration source. This analysis will provide insight into how often we need to re-compute the iEFC Jacobian on-sky. Finally, we aim to validate our tolerance model by comparing our simulation results to experimental iEFC data that we will be collecting on the Space Coronagraph Optical Bench (SCOOB) at the University of Arizona\cite{vangorkom2024}.

\acknowledgments 
 
This work is supported by NASA APRA grant \#80NSSC24K0288. This research made use of community-developed core Python packages, including: HCIPy\cite{por2018}, Astropy\cite{robitaille2013}, Matplotlib\cite{hunter2007}, and the IPython Interactive Computing architecture\cite{perez2007}. Joshua Liberman is a member of UCWAZ Local 7065.

\bibliography{report} 
\bibliographystyle{spiebib} 

\end{document}